# Enhanced optical performance of GaN Micro-light-emitting diodes with a single porous layer

Ziwen Yan, Xianfei Zhang, Yuyin Li, Zili Xie, Xiangqian Xiu, Dunjun Chen, Ping Han, Yi Shi, Rong Zhang, Youdou Zheng, and Peng Chen[*]

**AFFILIATIONS**

*Jiangsu Provincial Key Laboratory of Advanced Photonic and Electronic Materials and School of Electronic Science and Engineering, Nanjing University, Jiangsu, Nanjing, 210093, China.*
[*]**Authors to whom correspondence should be addressed:** pchen@nju.edu.cn

## Abstract

High-efficiency micro-light-emitting diodes (Micro-LEDs) are key devices for next-generation display technology. However, when the mesa size is reduced to around tens of micrometers or less, the luminous efficiency is constrained by the "efficiency-on-size effect". This work details the fabrication of gallium nitride (GaN) based Micro-LEDs with various mesa shapes and a single porous layer under the active region. A modified green LED epitaxial structure with different doped n-GaN layers combined with electrochemical etching created the porous layer. The strong light confinement achieved by the porous layer and the polygonal mesa greatly enhances spontaneous emission. The luminous intensity of the Micro-LEDs with the porous layer is approximately 22 times greater than those Micro-LEDs without the porous layer. A significant reduction in minimum full width at half maximum (FWHM) was observed in polygonal devices, suggesting a change in the luminescence mechanism. The influence of varying device geometry on emission performance was investigated. Experimental results reveal that, unlike circular porous Micro-LEDs, square and hexagonal porous Micro-LEDs exhibit more pronounced resonant emission, which provides a new technological approach for the further development of high-performance Micro-LEDs and lasers.

## Introduction

Owing to the fact that GaN material is a wide-bandgap semiconductor with good chemical stability and high electron mobility, its emission wavelength can be continuously adjusted from the ultraviolet to the visible light range, these make it widely used in fields such as full-color display, and optical communication[1-3]. GaN-based Micro-LED, as an extension of traditional GaN-based LED technology, possesses the inherent advantages of GaN-based LEDs and are solid-state light sources with high contrast, long lifespan, and high integration density[4-12]. With the rapid development of GaN-based Micro-LEDs and their increasing use high-resolution display, AR/VR, and medical fields, there is a growing demand for Micro-LED devices with high brightness and narrower linewidth[13-20]. Micro LEDs theoretically have high efficiency, but when the size is reduced to the order of tens of micrometers or less, their peak efficiency will significantly decrease. This phenomenon of efficiency decline due to size reduction is called the "efficiency-on-size effect". With the reduction of the Micro-LED chip size, the surface-to-volume ratio of the chip increases, and the non-radiative recombination induced by various unfavorable factors on each side surface increases. This is one of the main challenges in achieving high-efficiency small light-emitting chips.

In recent years, many research groups have addressed this challenge by adding resonant cavity structure [21-25]. Shuai Yang [26] et al. reported the fabrication of an aluminum mirror copper plate GaN-based green light resonant cavity LEDs (RCLEDs), which consist of an aluminum bottom mirror and two pairs of dielectric $TiO_2/SiO_2$ top distributed Bragg reflectors (DBRs), with the device emission wavelength of 518 nm and the FWHM of 14 nm. Some studies have found that when the GaN layer is transformed into a porous layer, the effective refractive index becomes smaller, and incorporating the porous layer into the DBRs structure is another way to achieve this[27-31]. Chuangjie Li et al. designed a GaN-based RCLED with a nanoporous GaN/n-GaN DBR and dielectric $SiO_2/Ta_2O_5$ DBRs, and the FWHM of the RCLED is 3.4 nm under the condition of electrical injection[32]. Chia-Feng Lin[33] et al. used a RCLED structure with the top and bottom porous GaN DBRs, the linewidth of the electroluminescence spectrum is reduced from 23.3 nm at 434 nm to 1.9 nm at 434.1 nm, due to the effect of the resonator effect. Although the aforementioned methods can address part of the issue, preparation techniques and precise epitaxial growth are often necessary for the fabrication of DBRs. At the same time, we believe that a single-layer porous GaN layer is easier to fabricate than a DBRs structure, and that the single-layer porous GaN layer is also beneficial for longitudinal current conduction.

In this study, we presented a green GaN based Micro-LED device featuring a single porous layer. A highly doped n-GaN layer is placed below the MWQs (InGaN/GaN) layer, and the highly doped n-GaN layer is transformed into a porous n-GaN layer by electrochemistry, which better confines the photons to the active region. The highest luminous intensity of the porous Micro-LED is 22 times higher than that of the conventional Micro-LED, and the minimum FWMH is approximately 5.9 nm. Three different shapes of single porous layer Micro-LEDs are designed, compared with the circular porous Micro-LEDs, the square and hexagonal porous Micro-LEDs exhibit more pronounced resonance emission.

## Material and Methods

### Structure and morphology of porous Micro-LED

Micro-LED is fabricated from a modified GaN-based green LED wafer grown on (111) Si substrate. As shown in Figure 1 (a), Upon an AlN/AlGaN buffer layer, the Micro-LED epitaxial structure is grown, and firstly followed by 1 μm thick undoped GaN (u-GaN). Subsequently a 1 μm lightly doped n-GaN ([Si]: $1.5 \times 10^{18}$/cm$^3$) layer and a 1 μm highly doped n-GaN([Si]:$8.5 \times 10^{18}$/cm$^3$) layer are grown，the highly doped n- GaN layer is then converted into a porous layer through electrochemistry, while the low doped n-GaN layer is used to conduct the circuit outside the electrolytic cell. To prevent the active region from being damaged during the etching process, a 100 nm u-GaN layer is grown between the n-GaN and the 150 nm InGaN/GaN MQWs. At the top of the wafer structure is a 100 nm thick p-GaN layer.

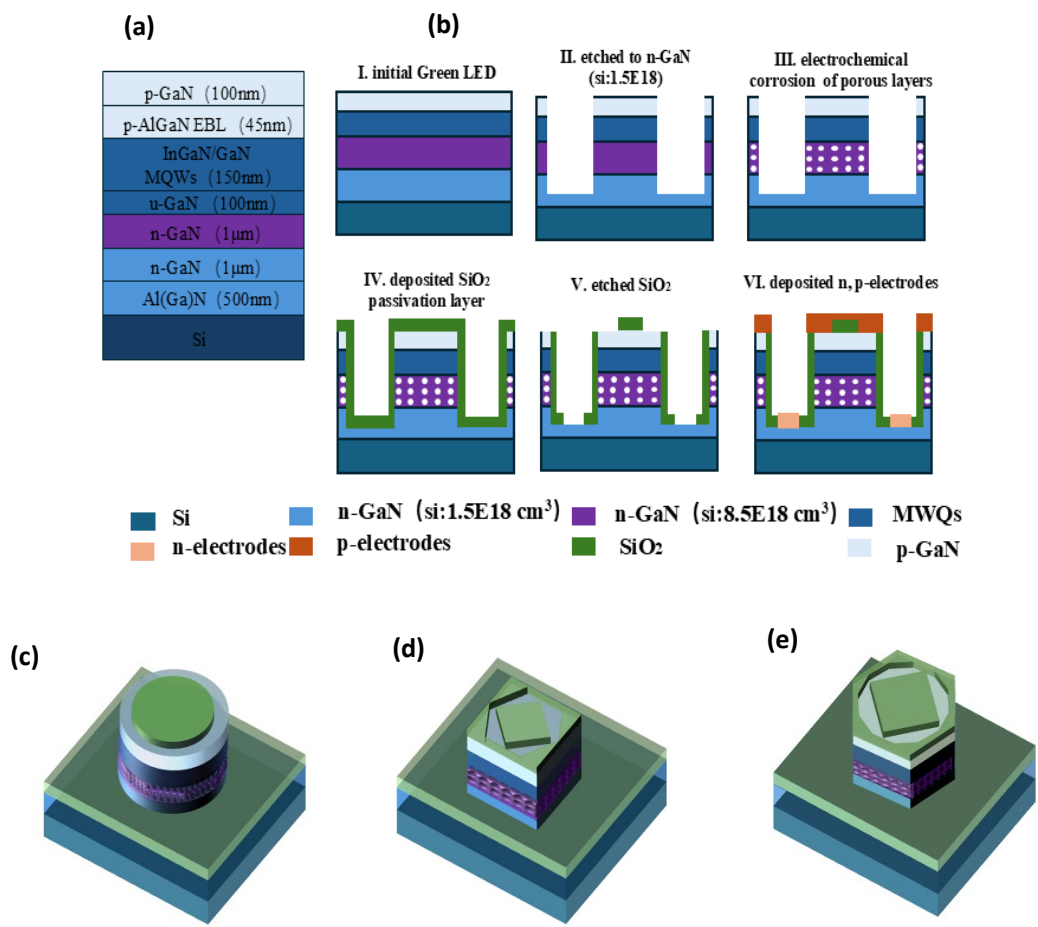

**Figure 1.** **The detailed fabrication process flow of the porous Micro-LEDs.** (**a**) Schematic diagram of the wafer structure. (**b**) schematic process flow for the formation of porous Micro-LEDs. **Schematic structure of the porous Micro-LEDs.** (**c**) circular Micro-LED structure. (**d**) square Micro-LED structure. (**e**) hexagonal Micro-LED structure.

The detailed process flow for the fabrication of the porous Micro-LED is depicted in Figure 1

(b). Initially, a silicon dioxide layer is deposited using plasma-enhanced chemical vapor deposition (PECVD) for pattern transfer. The Micro-LED mesa is formed after transferring the pattern to the GaN based LED wafer during the inductively coupled plasma (ICP) etch, which etches down to the low doped n-GaN layer. Subsequently, electrochemistry is performed to obtain a porous layer. Following, the buffered oxide etching (BOE) solution is used to remove $SiO_2$ from the sample's surface. Then $SiO_2$ is grown as a passivation layer using PECVD. After photolithography and $SiO_2$ etching expose the Micro-LED, the Cr/Al/Ni/Au layer is directly deposited on both the p-contact and the n-contact.

All the characterizations of Micro-LEDs are performed at room temperature. The morphology of the Micro-LEDs is observed by using optical microscope. The electroluminescence (EL) spectrum is measured using a Renishaw inVia Reflex micro-photoluminance spectroscopy system. The mesa photos and lighting photos are shown in the Figure 2. The current-voltage (I-V) characteristics are measured by direct-current power supply (Keithley 2601A).

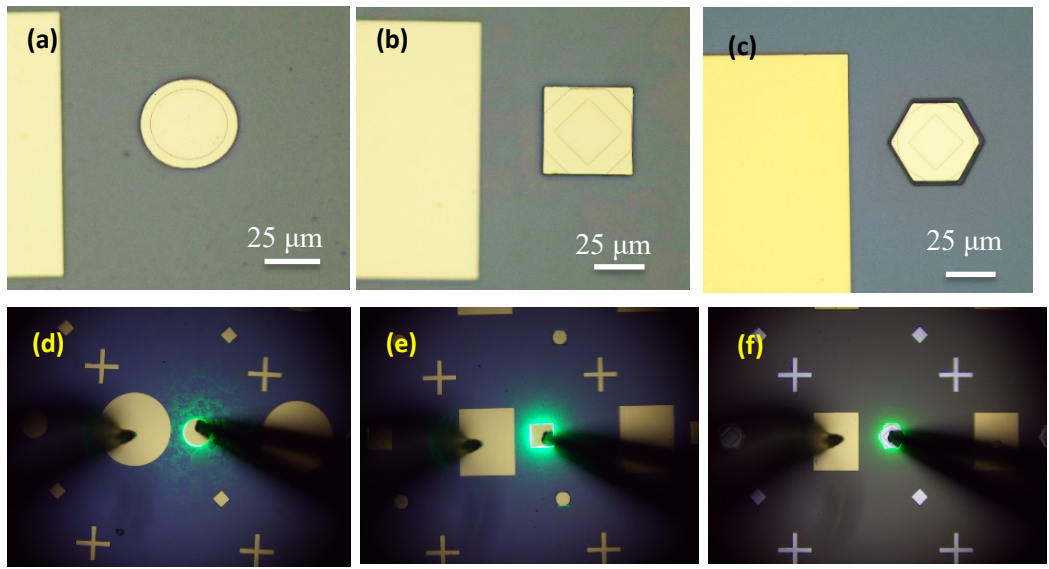

**Figure 2.** The optical microscope image of the Micro-LED. (**a**) circular Micro-LED. (**b**) square Micro-LED. (**c**) hexagonal Micro-LED. **Top-view luminous images of Micro-LED at 0.008 mA**. (**d**) circular Micro-LED. (**e**) square Micro-LED. (**f**) hexagonal Micro-LED.

## Results and Discussion

The existence of the porous layer fills the middle of the device with air, which will have a certain impact on the device's effective refractive index and resistance. The influence of the porous layer on the device's effective refractive index has been detailed in our previous research[34]. This study focuses primarily on the performance analysis of the porous Micro-LED under electrical injection, the influence of the porous layer on the electrical performance of the device is equally important. In this study, we calculate the effective resistance ratio of the porous Micro-LED. As we know resistance $R = \rho l / S$, where $l$ and $S$ are the length and cross-sectional area of the conductor, and $\rho$ is the resistivity. The porous layer can be defined as a series of air cylinders with spacing $d$ and diameter $D_{hole}$, the $l$ of the porous Micro-LED is the same as that of the normal Micro-LED, therefore:

$$r_S = S_{eff}/S_0 = (S_0 - S_{hole})/S_0 = 1 - n \cdot \pi r_{hole}^2/\pi r^2 \quad (1)$$

$r_S$ is the effective area ratio, $r$ is the radius of the Micro-LED, and $n$ is the number of pores,

$$n = \pi r^2 / (D_{hole}+d)^2 \quad (2)$$

So, the effective resistance ratio $r_R$:

$$r_R = R_{eff}/R_0 = S_0/S_{eff} = 1/r_S \quad (3)$$

The effective resistance ratio and effective refractive index of porous Micro-LED with different $D_{hole}$ shows in Figure 3 (a). When the $D_{hole}$ is 20, 28, 36, 42, 48, 54, 60 nm, the effective resistance ratio is 1.06, 1.12, 1.22, 1.33, 1.47, 1.69, 2.0. The effective resistance of the micro-LED increases with the increase of $D_{hole}$. When the $D_{hole}$ is the largest, the effective resistance doubles compared with the normal Micro-LED, so the increased resistance is still acceptable.

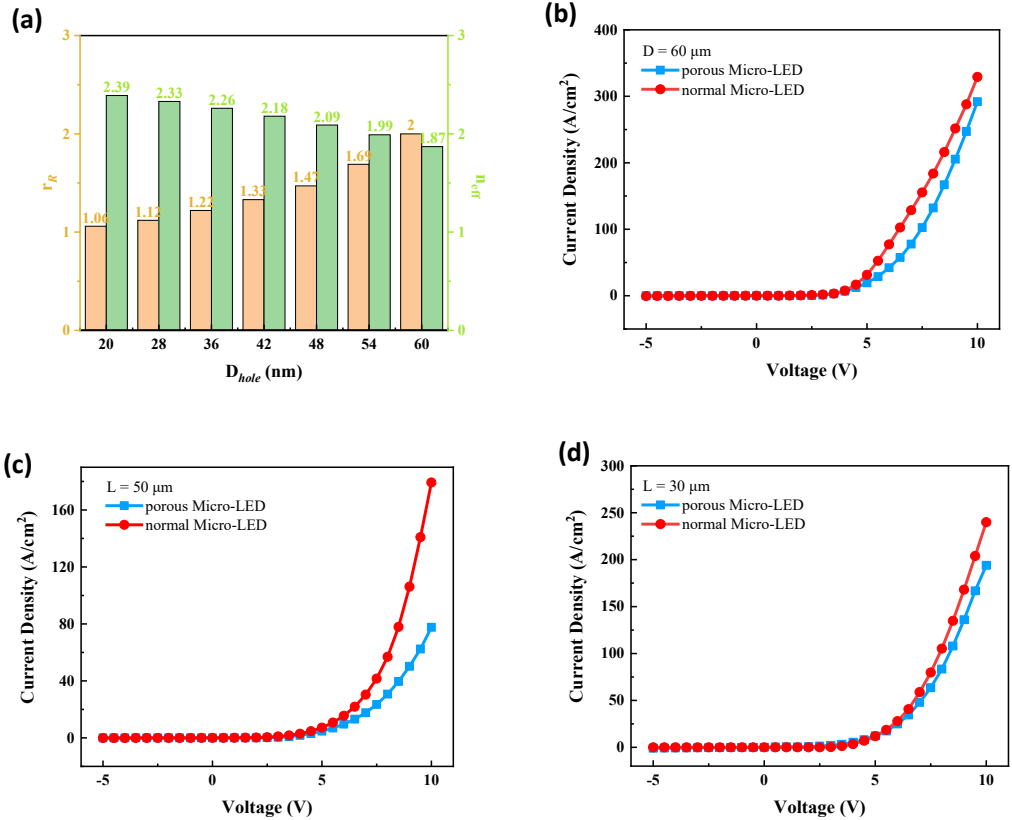

**Figure 3.** (**a**) Effective refractive index and resistance ratio of porous Micro-LED with different $D_{hole}$. (b-d) **J–V characteristic curves of different shapes of porous Micro-LEDs and normal Micro-LEDs** (**b**) circular Micro-LEDs with a diameter of 60 μm. (**c**) Square Micro-LEDs with a side length of 50 μm. (**d**) hexagonal Micro-LEDs with a side length of 25 μm.

As the aperture $D_{hole}$ increases, the effective resistance ratio of the Micro-LED also increases. This is well understood because, as the aperture continues to increase, more and more air enters the Micro-LED. The resistance of air is very high, hence the resistance of the porous Micro-LED increases correspondingly.

Figure 3 (b-d) illustrates the current density-voltage (J-V) characteristics of porous and normal Micro-LEDs with different shapes, respectively. The characteristics are obtained by applying a voltage range of -5 to 10 V to samples at room temperature. All samples demonstrate a sound turn-on voltage, a rapid increase in current density, indicating good device performance. Figure 3 (d) shows that when the hexagonal Micro-LED operates at a normal voltage of 7 V, the current densities of the porous and the normal Micro-LEDs are 34.5 A/cm$^2$ and 40.6 A/cm$^2$, respectively. In the case of the porous Micro-LED device, although the presence of the porous layer fills the device with air and results in a higher resistance, it does not significantly affect the device's current conduction.

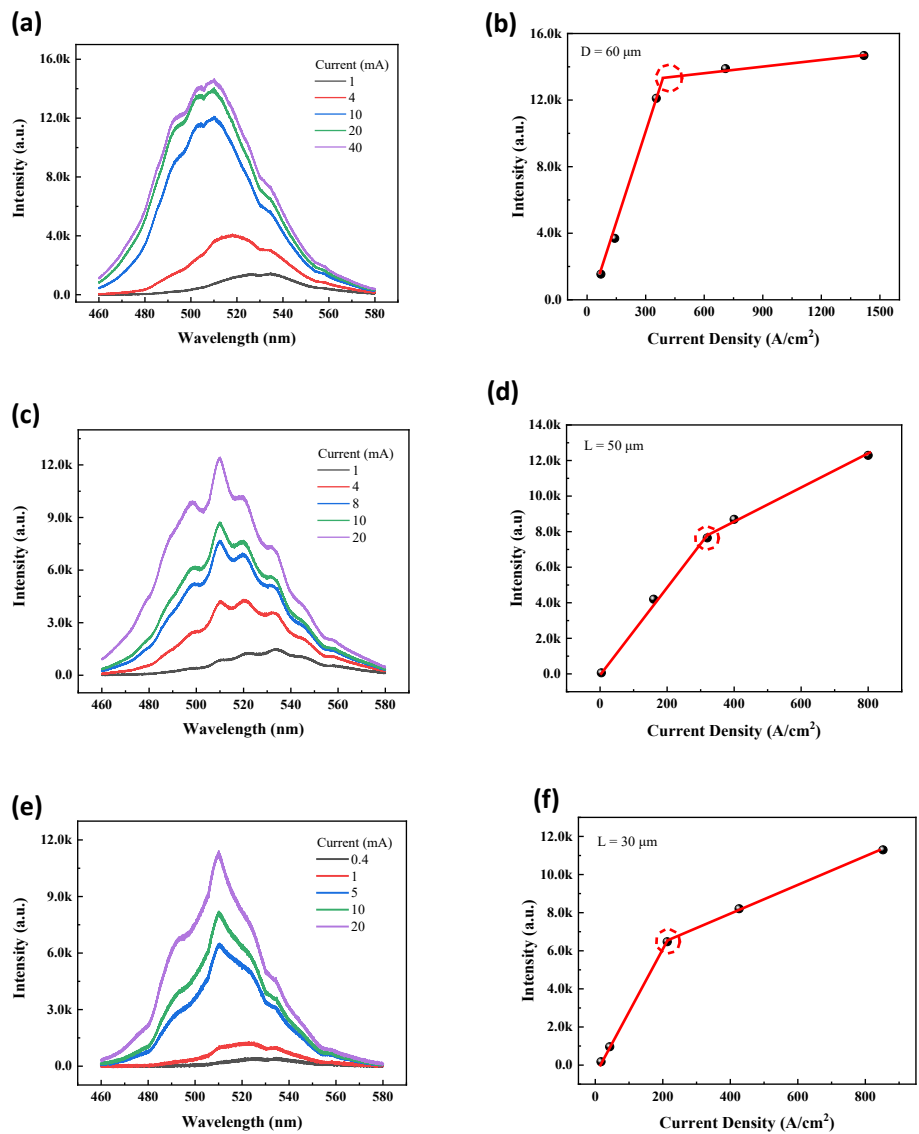

**Figure 4 Electroluminescence results of porous Micro-LEDs with different shapes** (**a**), (**c**), (**e**), EL spectra of porous Micro-LEDs with different shapes (**a**) circle, (**c**) square, (**e**) hexagon. (**b**), (**d**), (**f**) the main peak intensity as a function of current density (**b**) circular porous Micro-LED, (**d**) square porous Micro-LED, (**f**) hexagonal porous Micro-LED.

By varying the injection current, the EL emission spectra are measured for the circle, square, and hexagon porous Micro-LEDs, as shown in Figure 4 (a), (c), and(e), respectively. Under the low injection, the three different shapes of porous Micro-LEDs are dominated by spontaneous emission. As the injection current increases, the square porous Micro-LED gradually exhibits multiple resonance peaks, and the hexagonal porous Micro-LED displays a sharp peak at 510 nm with a gradually narrowing linewidth. However, the spectrum of the circular porous Micro-LED remains dominated by spontaneous emission, with a small spike at 510 nm appearing on the envelope of spontaneous emission. The change in device shape leads to different distributions of resonant modes and the light propagation distance within the resonant cavity, which are analyzed in detail in our previous work[35,36]. In the circular Micro-LED, the whispering gallery mode (WGM) is primarily near the edge of the device, which can be damaged due to the dry etching. In contrast, for the square and hexagonal porous Micro-LEDs, the mode is evenly distributed across the entire device. The resonance peaks, significantly different from the spontaneous emission, appear in the spectra of the square and hexagonal porous Micro-LEDs.

Figure 4 (b), (d), (f) display the main peak intensity of porous Micro-LEDs with different shapes. It can be found that the peak intensity of the three devices all exhibit a steep upward trend initially, then transition to a slower upward trend. For the circular porous Micro-LED, the inflection point of the peak intensity occurs between 354.6 A /cm$^2$ and 709.2 A /cm$^2$. For the square and hexagonal porous Micro-LEDs, the inflection points of peak intensity are 320 A/cm$^2$ and 213 A/cm$^2$, respectively. We attribute this phenomenon primarily to the device heating up due to the increase in current density, which causes the rising trend in device luminescence intensity to slow down.

On the same sample, half of the devices are corroded into pores by electrochemical etching, while the other half are not subjected to this process. Near the inflection points of the three devices, the EL spectral lines of the porous Micro-LEDs, with the same injection current, are selected and compared with those of the normal Micro-LEDs, as shown in Figure 5 (a), (c), (e). The luminous intensity of the porous Micro-LEDs with different shapes increases significantly. The luminous intensity of the circular, the square and the hexagonal porous Micro-LEDs is 2.64, 4.86 and 22 times that of the normal Micro-LEDs, respectively. When the injection current is 10 mA, the normal Micro-LEDs are dominated by spontaneous emission, while the porous Micro-LEDs not only increase the luminous intensity but also show a different peak from the spontaneous emission. This is mainly due to the existence of the porous layer, which increases the refractive index difference between the active region and the surrounding medium, thus enhancing the ability of photons to be confined in the active region and enhancing the resonance effect of the microcavity in the devices. Figure5 (b), (d), and (f) show the peak wavelength variation curves of porous and normal Micro-LEDs under different current densities. The peak wavelength of the porous Micro-LEDs shifts from 535 nm to 510 nm and then remains essentially unchanged at 510 nm. The peak wavelength of the normal Micro-LEDs also exhibits a blue shift, followed by continuous small amplitude changes. The foregoing results demonstrate that the porous layer significantly enhances micro-LED performance.

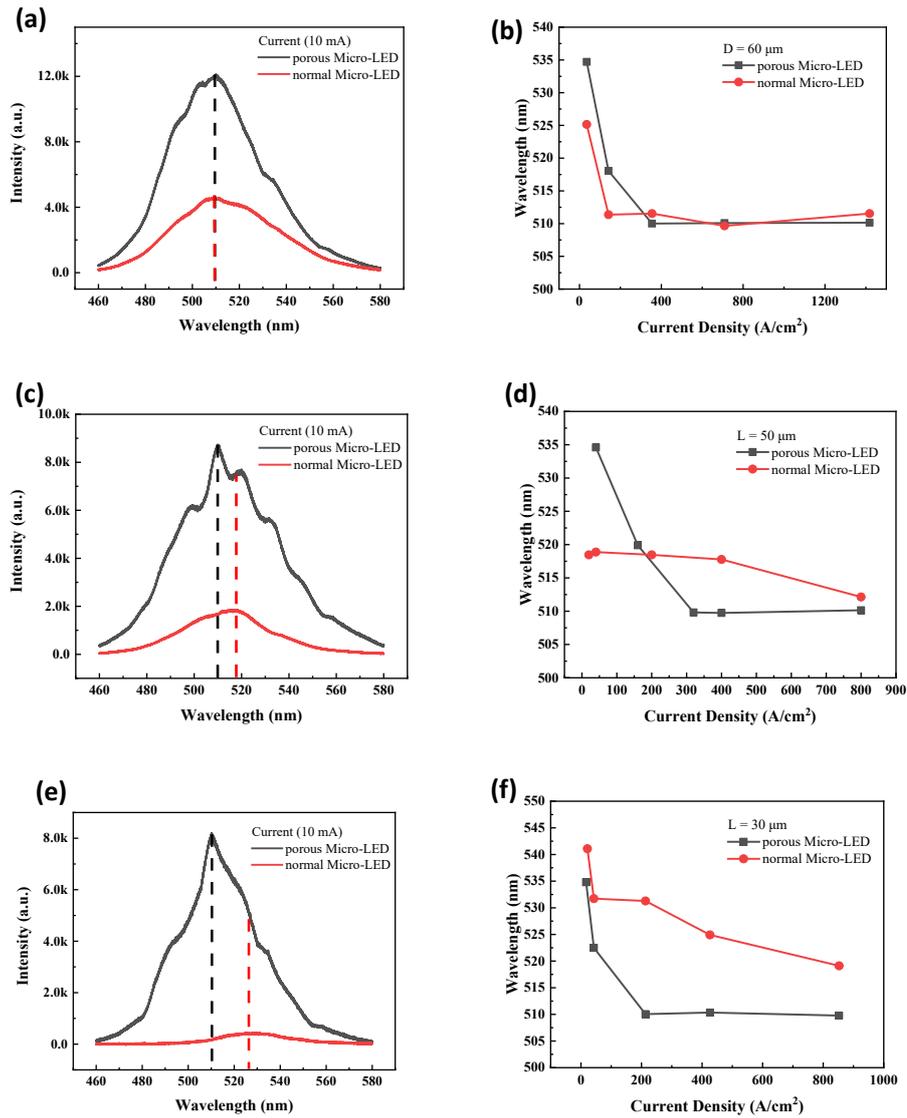

**Figure 5. EL spectra of porous and normal Micro-LED s with different shapes.** (**a**)circle, (**c**) square, (**e**) hexagon. **The peak wavelength of porous and normal Micro-LEDs with different shapes at different injection current levels.** (**b**)circle, (**d**) square, (**f**) hexagon

## Conclusion

In summary, we demonstrated GaN Micro-LEDs incorporating a single porous layer on a modified green LED epitaxial wafer. Electrochemical etching converted the highly doped n-GaN layer into a porous structure, effectively confining photons within the active region without compromising longitudinal current conduction. The luminous intensity of the Micro-LEDs with the porous layer is approximately 22 times greater than those Micro-LEDs without the porous layer. Variations in device geometry influenced resonance peak formation, with square and hexagonal micro-LEDs showing more pronounced resonant emission than circular counterparts, which provides a new technological approach for the further development of high-performance Micro-LEDs and lasers.

## FUNDING

This work is supported by National Natural Science Foundation of China (12074182), Collaborative Innovation Center of Solid-State Lighting and Energy-saving Electronics.

## AUTHOR DECLARATIONS

### Conflict of Interest

The authors have no conflicts to disclose.

### Author Contributions

**Yuyin Li:** Software (lead); Writing-original draft (equal); Investigation (equal). **Jing Zhou:** Writing-original draft (equal); Data curation (equal). **Ziwen Yan**: Formal analysis (equal). **Xianfei Zhang**: Investigation (equal). **Zili Xie**: Methodology (equal); Project administration (equal). **Xiangqian Xiu:** Resources (equal). **Dunjun Chen**: Methodology (equal). **Bin Liu**: Methodology (equal). **Hong Zhao**: Resources (equal); **Yi Shi**: Validation (equal). **Rong Zhang**: Supervision (equal). **Youdou Zheng**: Supervision (equal). **Peng Chen**: Conceptualization (lead); Funding acquisition (lead); Project administration (lead); Writing - review & editing (lead).

## DATA AVAILABILITY

The data that support the findings of this study are available from the corresponding authors upon reasonable request.